%% file: Mpara.tex
\documentclass[amsmath,amssymb,superscriptaddress,reprint]{revtex4-1}
\usepackage{graphicx}
\usepackage[dvipdfmx]{color}
\usepackage{dcolumn}
\usepackage{bm}
\usepackage{txfonts}
\usepackage{multirow}
\usepackage[utf8]{inputenc}
\usepackage[T1]{fontenc}
\usepackage{mathptmx}
\usepackage{etoolbox}

\makeatletter
\def\@email#1#2{%
 \endgroup
 \patchcmd{\titleblock@produce}
  {\frontmatter@RRAPformat}
  {\frontmatter@RRAPformat{\produce@RRAP{*#1\href{mailto:#2}{#2}}}\frontmatter@RRAPformat}
  {}{}
}%
\makeatother
\newcommand{\msr}{$\mu$SR}

\begin{document}
\title{Distinguishing Ion Dynamics from Muon Diffusion in Muon Spin Relaxation II -- Extension to Paramagnetic Muons}
\author{Ryosuke Kadono}\thanks{ryosuke.kadono@kek.jp}
\affiliation{Muon Science Laboratory, Institute of Materials Structure Science, High Energy Accelerator Research Organization (KEK), Tsukuba, Ibaraki 305-0801, Japan}
\author{Takashi U. Ito}
\affiliation{Advanced Science Research Center, Japan Atomic Energy Agency, Tokai-mura, Naka-gun, Ibaraki 319-1195, Japan}

\begin{abstract}%
We extend the previously published model that distinguishes between the diffusive motion of diamagnetic muons and the dynamics of ions around the muon in matter, and propose a generalized model for {\sl paramagnetic muons} (Mu$^0$s, bound states of a muon and an unpaired electron) observed in non-metallic host materials. The new model distinguishes among the independent motion of unpaired electron associated with Mu$^0$, the self-diffusive motion of Mu$^0$ as single atomic entity, and that of the ions surrounding Mu$^0$, where the muon spin relaxation is induced by dynamical fluctuations of the hyperfine (HF) field exerted from the unpaired electron (e.g., due to spin/charge exchange reaction) and/or that of the nuclear hyperfine (NHF) interactions between the unpaired electron and the surrounding ions. We have applied this model to the muonated radicals (Mu$^0$s in a polaron state) in conducting polymers, and examined the validity of the interpretations claimed in the earlier literature that the spin relaxation is induced by quasi-one dimensional jump motion of the unpaired electron. The result suggests that experimental support for such a claim is still inadequate and needs to be reexamined, including the possibility of other origins for the fluctuations. It is expected that our model will prove a useful guide for \msr\ studies of various local dynamics involving paramagnetic muon states.
\end{abstract}

\maketitle

\section{Introduction}
In nonmagnetic materials, implanted muons without accompanying unpaired electrons  (called diamagnetic muons)  exhibit spin relaxation mainly by sensing randomly distributed internal magnetic fields and their dynamical fluctuations exerted from the neighboring nuclear magnetic moments. It is established that the time evolution of muon spin polarization observed in such cases can be well described by the dynamical Kubo-Toyabe (KT) relaxation function \cite{Hayano:79,Yaouanc:10,Blundell:21}. 
Regarding the cause of fluctuations in the internal magnetic field, there are two general possibilities: one is due to the self-diffusion of muons, and the other due to the local dynamics of ions that carry nuclear magnetic moments.  While it has traditionally been believed that it is difficult to distinguish between these two cases, we have recently shown that it is possible by reexamining the assumptions underlying this function \cite{Ito:24}.

\begin{figure}[t]
  \centering
\includegraphics[width=0.95\linewidth]{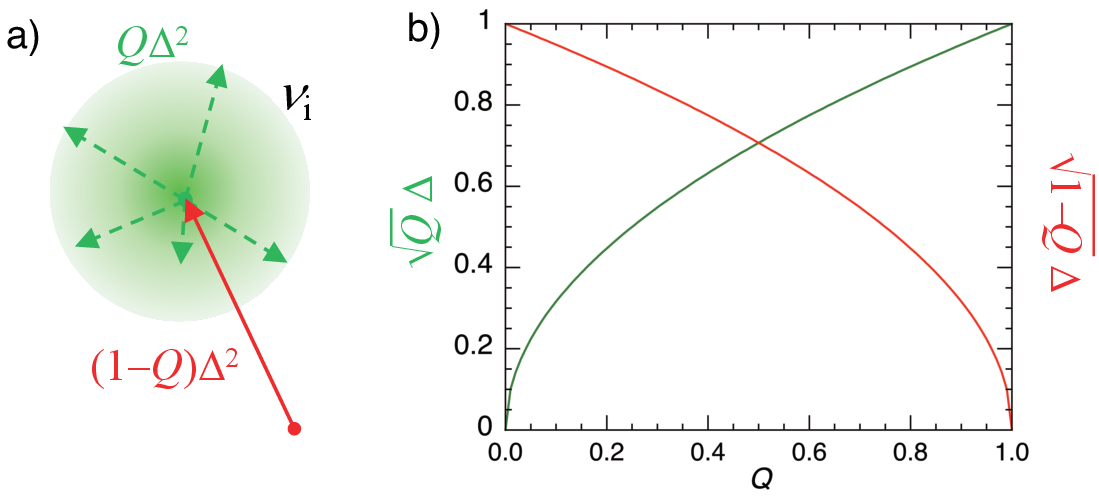}
\caption{(Color online) (a) Schematic of local fields consisting of static and fluctuating components, where $Q$ is the parameter representing the fractional amplitude of the local fields  fluctuating at a frequency $\nu_i$. (b) $Q$ dependence of the linewidths of the dynamical ($\sqrt{Q}\Delta$) and static ($\sqrt{1-Q}\Delta$) components corresponding to (a) (where $\Delta$ is set to unity).}
\label{EA}
\end{figure}

Specifically, we consider a generalized version of the autocorrelation function used to describe the fluctuations of the time-dependent internal magnetic field ${\bm H}(t)$ that considers the Edwards-Anderson parameter $Q$ $(0\le Q\le1$) \cite{Edwards:75,Edwards:76},
\begin{equation}
C(t)=\frac{\langle {\bm H}(0){\bm H}(t)\rangle}{\Delta^2/\gamma_\mu^2}\approx (1-Q)+Qe^{-\nu_{\rm i} t}, \label{Acf}
\end{equation}  
where $\langle...\rangle$ denotes the thermal average over the canonical ensemble, $\Delta$ is the linewidth of the spin relaxation determined by the second moment of the static field distribution $n({\bm H})$, $\gamma_\mu$ ($=2\pi\times 135.54$ MHz/T) is the muon gyromagnetic ratio, and $\nu_{\rm i}$ is the mean frequency of fluctuation: the meaning of the parameters is illustrated in Fig.~\ref{EA}(a). 
Here, the case $Q=1$ corresponds to the so-called strong collision model (also called random Markov process or random phase approximation) with which the dynamical KT function, $G_z^{\rm KT}(t)$,  is derived. In this case, it is implicitly assumed that the fluctuations of ${\bm H}(t)$ depend only on the relative change between the muon spin ${\bm S}_\mu(t)$ and ${\bm H}(t)$, so it is impossible within the model to distinguish which one is actually in motion.

On the other hand, when the fluctuations are caused by local ion dynamics, some fraction of the ions in proximity to the muon may be stationary. This means that the corresponding ${\bm H}(t)$ will have a residual component in the autocorrelation before and after the change, corresponding to $Q<1$ (hereinafter called `partial correlation model'). As shown in in Fig.~\ref{EA}(b), $Q$ defines the effective linewidth for the static and dynamical components in the relaxation function. It is important to note that when the fluctuations are due to local dynamics of ions, the parameter $1-Q$, which represents the fractional amplitude of the ions not in simultaneous motion, is expected to have a value much greater than zero and can be distinguished from the case where muons are diffusing (i.e., $Q=1$).  Hereafter, the fluctuation frequency for $Q=1$ will be referred to as $\nu$ to distinguish it from $\nu_{\rm i}$ for $Q<1$.

The spin relaxation function for the partial correlation model, $G_z^{\rm ID}(t)$, was derived by Monte Carlo simulation for various values of $Q$, and revealed to exhibit qualitatively different behavior from $G_z^{\rm KT}(t)$ in terms of fluctuation rate dependence \cite{Ito:24}. To examine the validity of the partial correlation model, $G_z^{\rm ID}(t)$ was used to re-analyze data obtained from \msr\ measurements of the hybrid perovskite FAPbI$_3$ (FA denotes the formamidinium molecule HC(NH$_2$)$_2$) \cite{Ito:24}. Here, the random rotational motion of FA molecules in the perovskite lattice causes partial fluctuations in ${\bm H}(t)$. While the previous analysis using $G_z^{\rm KT}(t)$ suggested that the linewidth $\Delta$ decreases significantly as the temperature rises and that $\nu$ exhibits unexpectedly small values with a strange peak in a narrow temperature range \cite{Koda:22}, $G_z^{\rm ID}(t)$ is successful to account for the overall  temperature variation of the spectra with $(1-Q)^{1/2}\simeq0.2$, and $\nu_{\rm i}$ obtained in the analysis shows a monotonous increase with temperature as expected. 

Based on this success, we extend the applicability of this partial correlation model to  construct a microscopic model for the longitudinal spin relaxation exhibited by muons in the paramagnetic state, where a single unpaired electron undergoes hyperfine (HF) interactions with both muon and surrounding nuclear spins (with the latter specifically called nuclear hyperfine (NHF) interaction).  In such situation, the muon spin relaxation is dominated by the thermal fluctuation (spin relaxation) of the unpaired electron. It is then naturally assumed in terms of mass that the electron is the most mobile, followed by the muon, and finally the surrounding atoms. Since the strong collision model cannot distinguish between these sources, we again resort to the partial correlation model where two characteristic parameters for the \msr\ time spectra, the initial polarization of the paramagnetic muon and spin relaxation rate versus longitudinal field (LF), serves as powerful identifiers for the dynamics. Our Monte Carlo simulation for the NHF interaction shows that they exhibit significant difference from those predicted by the strong collision model, and that a comparison with experiments is likely to distinguish the sources of fluctuations. Furthermore, a reexamination of the previous \msr\ studies in conducting polymers indicates that the earlier interpretation that the cause of spin relaxation is the motion of unpaired electrons (carriers) needs to be reconsidered.

\section{Electronic state of paramagnetic muons and their dynamics}
\subsection{The origin of paramagnetic Mu in matter}

When muons as positively charged ion beams ($\mu^+$s) are implanted into crystalline solids, they are localized at the interstitial site(s) as atomic defects.  Since the electric neutrality is maintained in the host material, electrons are also provided in pairs with the $\mu^+$s to compensate the positive charge. Therefore, the muon can mimic the local electronic structure of hydrogen as a light isotope (pseudo-H), which is denoted below by the element symbol Mu.

Provided that the interstitial Mu is in thermal equilibrium in the host crystal, it is predicted in many cases that Mu$^0$ cannot exist stably due to electron-phonon interactions, as has been established for the case of interstitial hydrogen (H$_{\rm i}$) in semiconductors \cite{Pankove:91,Pearton:92,Walle:03}. This can be qualitatively understood by a simple Hamiltonian that introduced by Anderson,
\begin{equation}
V=-\eta x(n_\uparrow+n_\downarrow)+\frac{1}{2}cx^2,
\end{equation}
where the first term is a linear electron-phonon coupling term for the displacement $x$ of the bond length between the dopant H$_{\rm i}$ and host atoms with $n_\uparrow$ and $n_\downarrow$ denoting the occupancies (0 or 1) of the spin-up and spin-down bond orbitals, and the second term is a quadratic elastic restoring term \cite{Anderson:75,Watkins:84,Coutinho:20}.  The energy difference between the case of 2H$_{\rm i}^0$ ($-\eta^2/2c \times2$) and H$_{\rm i}^-$ ($-2\eta^2/c$) is $-\eta^2/c$. Adding the Hubbard correlation energy $U$, the net correlation energy associated with the energy difference between 2H$_{\rm i}^0$ and the disproportionated state (${\rm H}_{\rm i}^+ + {\rm H}_{\rm i}^-$) becomes $U_{\rm eff}=U-\eta^2/c$. Therefore, if $\eta$ is sufficiently large to yield $U_{\rm eff}<0$, H$_{\rm i}^-$ is energetically most favorable, and it behaves as a `negative-$U$' center in thermal equilibrium. Here, in the context of our discussion, one H$_{\rm i}$ represents the muon and the other represents the host.

In reality, however, Mu$^0$s or muonated radicals (those in a polaronic state, Mu$\dot{R}$) are often observed upon $\mu^+$'s implantation to non-metallic solids. In these cases, it has been shown experimentally that the electrons are provided by excitations of anions (valence band electrons) due to the kinetic energy of the muon itself (typically $\sim$4 MeV). In particular, if the bottom of the conduction band consists of $s$ orbitals of cation atoms, the cross section for the excited electrons to be captured by localized muons can be sufficiently large due to the large electron mobility. In molecular materials with unsaturated bonds, similar electronic excitation will cause an addition reaction for Mu by cleaving double bonds to undergo HF interactions from the remaining unpaired electron localized at the adjacent ions. Note that such electronic excitation also occurs with ultraviolet light or x-ray irradiation, which is known to generate paramagnetic hydrogen (H$^0$) in ionic crystals containing trace amounts of H.  The behavior of H/Mu under electronic excitation has recently been established as a manifestation of the ambipolar nature of hydrogen, where H$^0$/Mu$^0$ is understood as a donor/acceptor-like state \cite{Hiraishi:22,Kadono:24}. Since Mu$\dot{R}$ can be regarded as a subclass (polaron state) of Mu$^0$, the paramagnetic Mu is represented by the latter in the following.

Considering that the Mu$^0$ states cannot exist stably in thermal equilibrium, these states are presumed to exist as metastable states (= relaxed excited states) whose lifetime is longer than the average lifetime of muons \cite{Hiraishi:22,Kadono:24}. It should then reminded that the same is true for the diamagnetic states (Mu$^+$ or Mu$^-$) in terms of being under electronic excitation, and that it would not be trivial whether such states correspond to those of H in thermal equilibrium. As we will see below, when Mu$^0$ undergoes a fast spin/charge exchange reaction with the carriers excited in the conduction band, the response of Mu$^0$ to the magnetic field approaches that of the diamagnetic Mu (the `motional narrowing' effect).

In any case, the advantage of using Mu$^0$ as a microscopic probe of matter is that it exhibits an effective gyromagnetic ratio that is two orders of magnitude larger than that of diamagnetic Mu, making it highly sensitive to the internal magnetic field. For example, at low magnetic fields, the effective gyromagnetic ratio becomes
\begin{equation}
\gamma_{\rm av} = \frac{\gamma_\mu+\gamma_{\rm e}}{2}=2\pi\times14079.87\:\:{\rm MHz/T},
\end{equation}
which is 104 times higher than $\gamma_\mu$ due to the large electron gyromagnetic ratio ($\gamma_{\rm e}=2\pi\times 28024.21$ MHz/T). Therefore, besides the studies on the local electronic states of Mu as pseudo-H, Mu$^0$ can be used as a sensitive probe for obtaining microscopic information on the bulk properties of host materials such as dielectrics (which is difficult using diamagnetic Mu). In practice, however, the studies on Mu$^0$ or Mu$\dot{R}$ that have been performed so far have focused mainly on the local electronic structure as pseudo-H based on spectroscopic techniques, and a limited example of the latter is the study of carrier dynamics in conducting polymers.

For example, linear-conjugated polymers with alternating single and double carbon bonds can be regarded as one-dimensional (1D) metals with a band gap due to bond alternation (the Peierls transition). Therefore, if carriers can be doped into the conduction band by adding impurities to these polymers, they may exhibit electrical conductivity; this has been experimentally confirmed for several polymers including polyacetylene (PA) as a prototype \cite{Shirakawa:71,Su:79}. In the case of muon implantation to these polymers, it is expected that the unpaired electrons produced with Mu$\dot{R}$ addition may enter the conduction band (corresponding to the lowest unoccupied molecular orbital) to form polarons (or `solitons' in {\it trans}-PA), and they undergo charge exchange reaction with Mu during the quasi-1D jump motion, ${\rm Mu}\dot{R}\rightleftarrows{\rm Mu}^+ + e^-$. In fact, previous $\mu$SR results in {\it trans}-PA were interpreted by this scenario \cite{Nagamine:84,Ishida:85}, followed by similar \msr\ studies for conjugated polymers such as polyaniline (PANI) \cite{Pratt:97} and poly(3-hexylthiophene) (P3HT) \cite{Risdiana:10a,Risdiana:10b}.

\subsection{$\mu$SR time spectra for the paramagnetic Mu}

\begin{figure}[t]
  \centering
\includegraphics[width=0.5\linewidth]{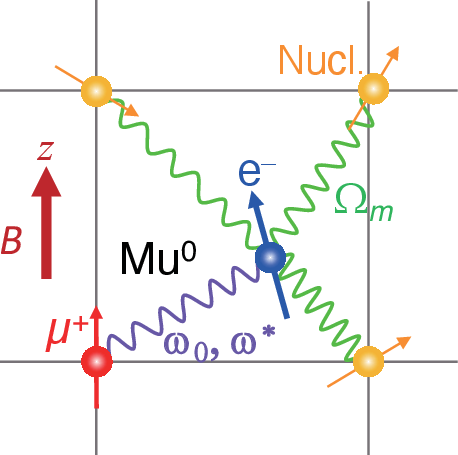}
\caption{(Color online)  Schematic of paramagnetic Mu state: an unpaired electron undergoes HF interaction with muon ($\omega_0$, $\omega^*$) and NHF interactions with surrounding nuclei ($\Omega_m$). }
\label{Mup}
\end{figure}

In general, a muon in the paramagnetic state is subject to HF interaction with an unpaired electron, and the latter to NHF interaction with the surrounding nuclear spins, which is schematically shown in Fig.~\ref{Mup}.  Their electronic states are described by the following Hamiltonian,
\begin{equation}
\mathcal{H}/\hbar = 
\frac{1}{4}\omega_0{\bm \sigma}\cdot{\bf \tau} +\frac{1}{4}\omega_*({\bm \sigma}\cdot{\bm n})({\bm \tau}\cdot{\bm n}) -\frac{1}{2}\omega_\mu\sigma_z + \frac{1}{2}\omega_e\tau_z\nonumber 
\end{equation}\vspace{-6mm}
\begin{equation}
+\sum_m[\Omega_{\perp_m}{\bm S}_e\cdot{\bm I}_m +(\Omega_{\parallel_m}-\Omega_{\perp_m})S_e^zI_m^z]- \sum_m\omega_{n_m}I_m^z,\label{hn}
\end{equation}
where $\omega_0$ and $\omega_*$ respectively denote the transverse ($\omega_\perp$) and anisotropic ($\omega_\parallel-\omega_\perp$) parts of the HF interaction, $\omega_\mu=\gamma_\mu B$, $\omega_e=\gamma_e B$, ${\bm \sigma}$ and ${\bm \tau}$ are the Pauli spin operators for muon and electron, and ${\bm n}$ is a unit vector along the symmetry axis \cite{Percival:79,Patterson:88}, $\Omega_{\perp_m}$ and $\Omega_{\parallel_m}-\Omega_{\perp_m}$ are the transverse and anisotropic parts of the NHF parameter for the $m$th nuclear spin $I_m$ ($=\frac{1}{2}$ for $^1$H), $\omega_{n_m}=\gamma_{n_m}B$ with $\gamma_{n_m}$ being the gyromagnetic ratio for the corresponding nuclei; $\Omega_{\parallel_m}$ and $\Omega_{\perp_m}$ are usually dominated by the Fermi contact interaction with an order of magnitude smaller contributions from magnetic dipolar interaction.

In the experiment to study the dynamics of Mu$^0$, an external magnetic field $B$ (in parallel with the initial muon polarization, or LF) is applied, and the time evolution of the spin polarization $P_z(t)$ (longitudinal spin relaxation) is measured. In the following, the behavior of $P_z(t)$ is discussed for two cases, one in which the fluctuations of ${\bm H}(t)$ are represented by the strong collision model  ($Q=1$ and $\nu_{\rm i}=\nu$ in Eq.~(\ref{Acf})), and another where ${\bm H}(t)$ is described by the partial correlation model ($0<Q<1$), paying attention to the assumed physical situations for the respective cases.

\subsubsection{Static case ($\nu=\nu_{\rm i}=0$)}
When all interactions described by Eq.~(\ref{hn}) are static, there is no longitudinal spin relaxation (i.e.~the corresponding rate $1/T_{1\mu} = 0$). The spin rotation induced by the transverse component of the HF/NHF fields relaxes rapidly due to the random distribution of the NHF fields \cite{Patterson:88}.  Consequently, the experimentally observable $P_z(t)$ for the isotropic Mu$^0$ ($\omega_0\gg\omega^*$) is given by the time-independent initial polarization as follows
\begin{equation}
P_z(t;x_{\rm p}) \simeq g_z(x_{\rm p}),\label{Alf}  %
\end{equation}
\begin{eqnarray}
g_z(x_{\rm p})&=& \frac{\frac{1}{2}h_z(x_{\rm n})+x_{\rm p}^2}{1+x_{\rm p}^2},\:\:x_{\rm p}=2\gamma_{\rm av}B/\omega_0\label{gzp}\\
h_z(x_{\rm n})&\simeq& \frac{\frac{1}{3}+x_{\rm n}^2}{1+x_{\rm n}^2},\:\:x_{\rm n}\simeq\gamma_{\rm av}B/\Delta_{\rm n},
\end{eqnarray}
where $g_z(x_{\rm p})$ is the initial polarization of Mu$^0$ as a function of the normalized field $x_{\rm p}$ \cite{Patterson:88}, $h_z(x_{\rm n})$ is the approximated field dependence of initial polarization under the NHF interaction characterized by the linewidth $\Delta_{\rm n}$ (determined by the second moment of the distribution of NHF fields) with the $\frac{1}{3}$ term corresponding to the possibility that the direction of the NHF field is parallel with the initial spin polarization of the triplet Mu$^0$ state \cite{Beck:75}. This approximation is also supported by the fact that the spin relaxation of the triplet Mu$^0$ in solid Kr is well represented by the quasi-static KT relaxation function $G_z^{\rm KT}(t; \Delta_{\rm n},0)$ with $\gamma_\mu B$ replaced by $\gamma_{\rm av}B$, where a relatively small $\Delta_{\rm n}$ ($\approx0.7$ MHz) due to dilute $^{83}$Kr nuclei allows us to observe the detailed time evolution of $P_z(t)$  \cite{Storchak:96}. In the usual experimental condition with relatively large $\Delta_{\rm n}$ ($\approx10^2$ MHz), only the asymptotic behavior, $h_z(x_{\rm n})\approx G_z^{\rm KT}(\Delta_{\rm n}t\gg1)$, can be observed. (The exact form of $h_z(x_{\rm n})$ under LF is found elsewhere \cite{Yaouanc:10}.)   The NHF parameter in the static limit can be written more explicitly using the parameters in Eq.~(\ref{hn}) \cite{Kadono:90},
\begin{equation}
\Delta_{\rm n}^2=\sum_m\frac{1}{3}(\Omega_{\perp_m}^2+2\Omega_{\parallel_m}^2)I_m(I_m+1).
\end{equation}

\begin{figure}[t]
  \centering
\includegraphics[width=0.6\linewidth]{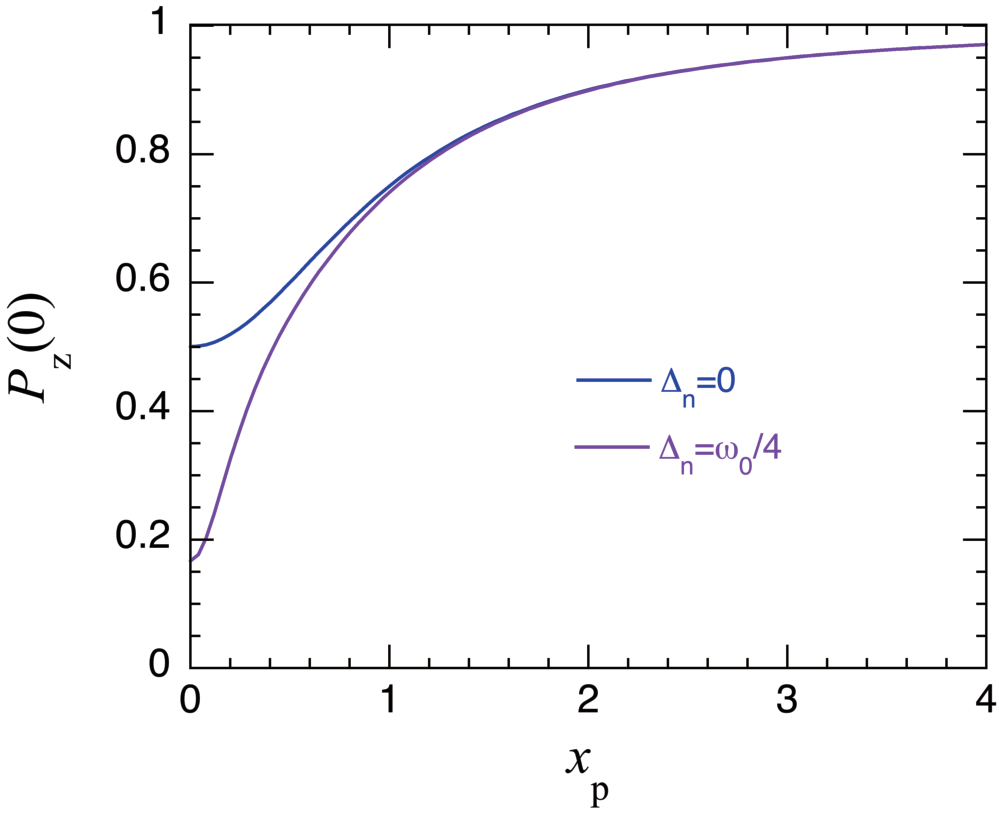}
\caption{(Color online)  LF ($x_{\rm p}$) dependence of the initial muon polarization $P_z(t)$ for the NHF interaction $\Delta_{\rm n}=0$ (a) and $\Delta_{\rm n}=\omega_0/4$ (b). }
\label{Pz0n0}
\end{figure}

When the NHF interactions are negligible ($\Delta_{\rm n}=0$), we have $h_z(x_{\rm n})=1$ and 
\begin{eqnarray}
P_z(t;x_{\rm p})&=& g_z(x_{\rm p})+[1-g_z(x_{\rm p})]\cos\omega_{24}t,\label{Alf0} \\ %
\omega_{24}&=&\omega_0(1+x_{\rm p}^2)^{1/2}\nonumber  %
\end{eqnarray}
where $\omega_{24}$ is the transition frequency between spin triplet- and singlet-Mu$^0$ states. Practically, it is difficult to observe the rotation signal for $\omega_{24}$ under usual experimental conditions, and Eq.~(\ref{Alf}) representing the time average of Eq.~(\ref{Alf0}) is actually observed.
An example of the initial polarization $P_z(0)$ vs.~LF is shown in Fig.~\ref{Pz0n0} for the case of $\Delta_{\rm n}=\omega_0/4$ together with that without NHF interaction. In the case of Mu$\dot{R}$s in organic materials, the NHF interaction is complicated by the anisotropic electronic structure, and $P_z(0)$ is obtained by numerical calculations combined with {\sl ab initio} density functional theory calculations to estimate Mu site(s) \cite{Pratt:22}.

\subsubsection{Spin dynamics treated by the strong collision model ($Q=1$)}

\begin{figure}[t]
  \centering
\includegraphics[width=0.95\linewidth]{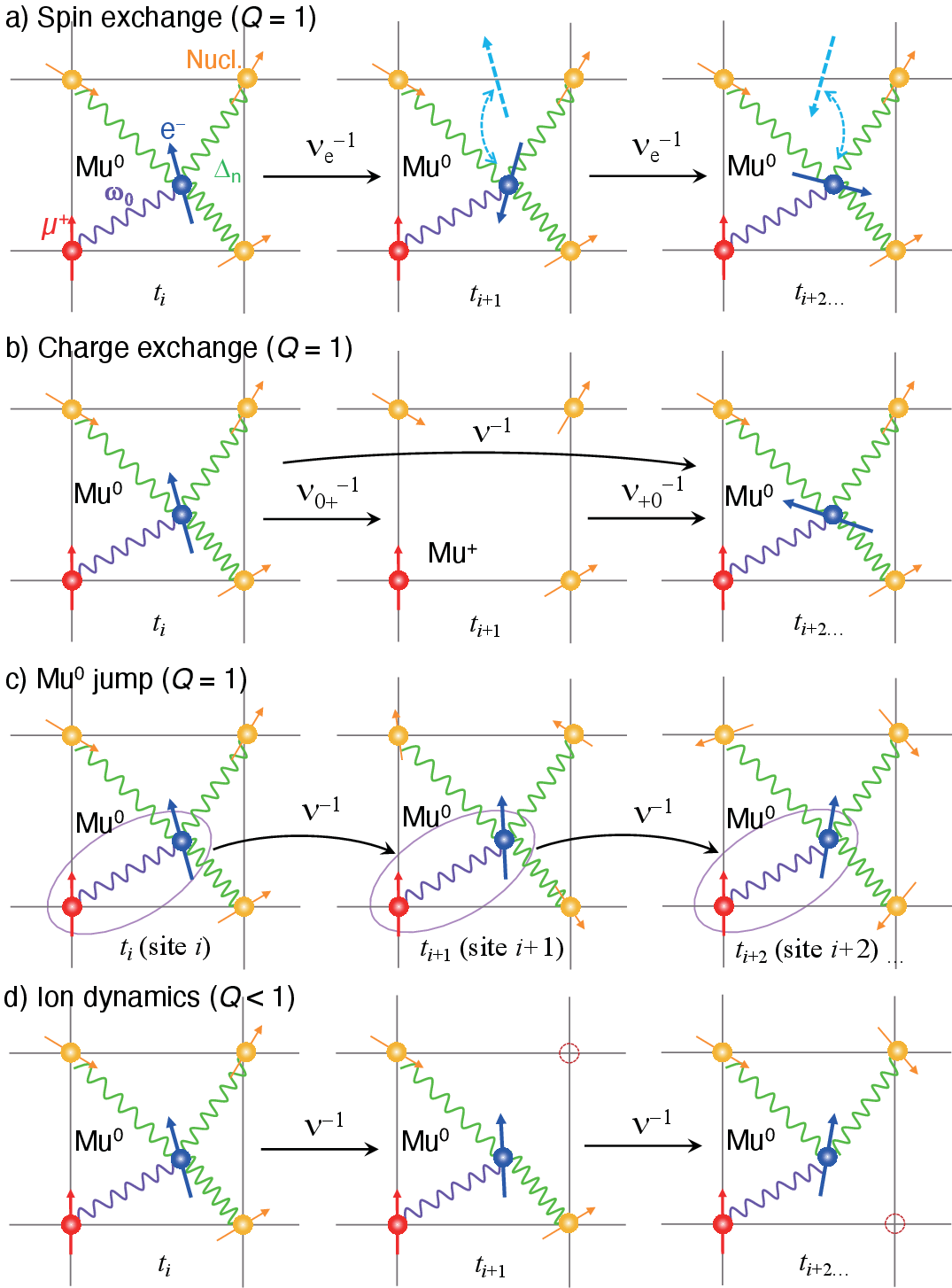}
\caption{(Color online)  Schematic of time sequences expected for the paramagnetic Mu state: (a) spin exchange with the mean frequency $\nu_e$, (b) charge change with the mean frequency $\nu=\nu_{0+}+\nu_{+0}$ (where $\nu_{0+}$ and $\nu_{+0}$ are respectively the electron release and capture rates), (c) Mu$^0$ jump motion, and (d) ion dynamics with the mean frequency $\nu$. In (a) and (b), $\omega_0$ and $\Delta_{\rm n}$ are fully reset at every time sequence ($t_i \rightarrow t_{i+1}\rightarrow t_{i+2} \rightarrow$...), re-initializing the evolution of muon polarization.  (c) $\Delta_{\rm n}$ is reset while $\omega_0$ is unchanged (keeping an identity as Mu$^0$). (d) Only a part of $\Delta_{\rm n}$ (and/or $\omega_0$) fluctuates.}
\label{dymod}
\end{figure}

When the magnetic field probed by Mu$^0$ fluctuates, there are several possible causes as inferred from Fig.~\ref{Mup}. Among them, there are two cases where the strong collision model (corresponding to $Q=1$) is appropriate to evaluate the behavior of $P_z(t)$, namely, (i) the HF field ($\omega_0$) fluctuates due to a direct state change of unpaired electron (spin and/or charge exchange reactions): such a situation is often suggested for the dynamics of Mu$\dot{R}$ attached to isolated molecules in the gas phase or in solution \cite{Fleming:96,McKenzie:11}, and (ii) the NHF field ($\Delta_{\rm n}$) fluctuates due to diffusive (jumping) motion of Mu$^0$ itself \cite{Kiefl:89,Kadono:90,Kadono:91,Kadono:94,Gxawu:05}: schematic illustrations are found in Fig.~\ref{dymod}.  The interpretation of the experimental results for conducting polymers argued in the earlier literature corresponds to the case of Fig.~\ref{dymod}(b) \cite{Nagamine:84,Ishida:85,Pratt:97,Blundell:02,Pratt:04,Risdiana:10a,Risdiana:10b}. 

\subsubsection*{2.1 Spin and/or charge exchange interaction}

\begin{figure}[t]
  \centering
\includegraphics[width=0.95\linewidth]{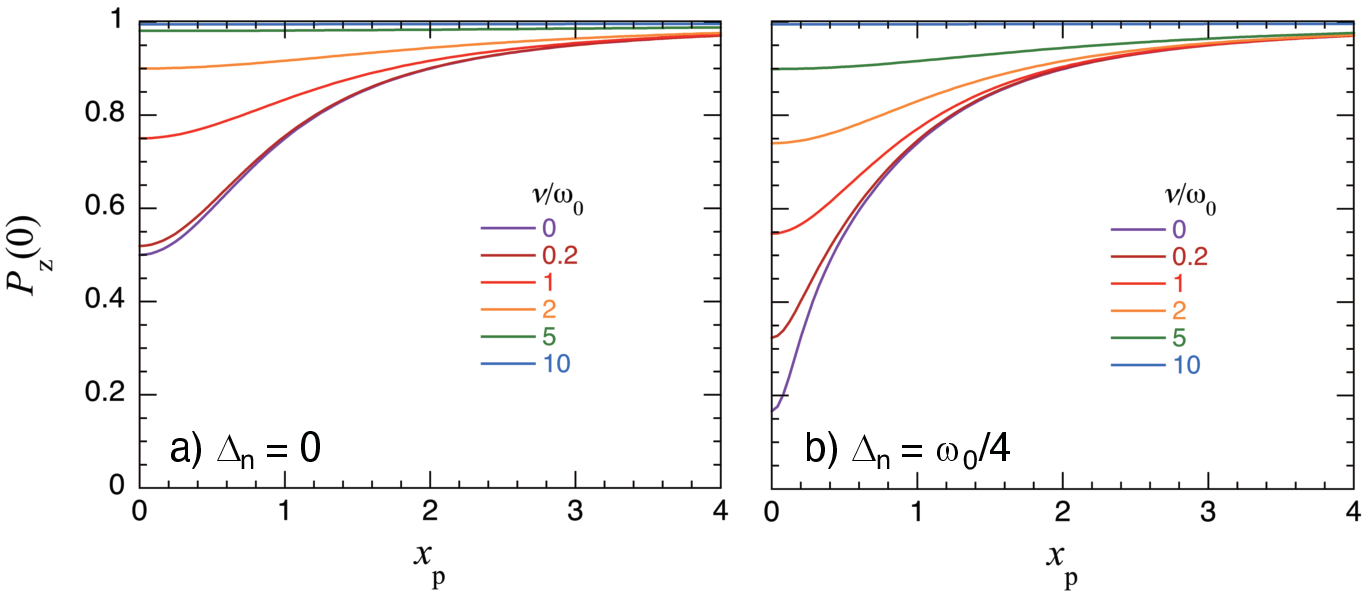}
\caption{(Color online)  LF ($x_{\rm p}$) dependence of the initial muon polarization $P_z(t)$ when the HF field ($\omega_0$ fluctuates with the rate $\nu$: for the NHF interaction $\Delta_{\rm n}=0$ (a) and $\Delta_{\rm n}=\omega_0/4$ (b). $P_z(0)\rightarrow1$ with $\nu\gg\omega_0$ in both cases. }
\label{Pz0n}
\end{figure}

Let us assume that the fluctuation (relaxation) rate $\nu_e$ of unpaired electron is determined by the spin exchange reaction with localized paramagnetic defects (Fig.~\ref{dymod}(a)), and that it is independent of the charge exchange reaction rate $\nu=\nu_{0+}+\nu_{+0}$ (where $\nu_{0+}$ and $\nu_{+0}$ are respectively the electron release and capture rates, with $\nu^{-1}$ providing the upper limit for the duration time $\nu_{0+}^{-1}$ of HF field, Fig.~\ref{dymod}(b)). Then the behavior of $P_z(t)$ depends intricately on these two parameters. When $\nu_e/\nu\ll1$, the external field $B$ serves to hold the spin polarization of the electron bound to Mu, where the phenomenological theory by Nosov and Yakovleva predicts that  $P_z(t)$ is approximated as follows \cite{Nosov:63,Nosov:65}:
\begin{equation}
P_z(t;x_{\rm p})\simeq g_z(x_{\rm p};\nu)\exp[-t/T_{1\mu}],\label{Alfd}  %
\end{equation}
\begin{eqnarray}
g_z(x_{\rm p};\nu)&\simeq& \frac{\frac{1}{2}h_z(x_{\rm n})+x_{\rm p}^2+\nu^2/\omega_0^2}{1+x_{\rm p}^2+\nu^2/\omega_0^2},\label{gzpd}\\
h_z(x_{\rm n};\nu)&\simeq& \frac{\frac{1}{3}+x_{\rm n}^2+\nu^2/\Delta_{\rm n}^2}{1+x_{\rm n}^2+\nu^2/\Delta_{\rm n}^2},\label{hzd}
\end{eqnarray}
where Eq.~(\ref{hzd}) is derived by the same sprit as Eq.~(\ref{gzpd}). The term $\nu^2/\omega_0^2$ is intuitively understood to come from the fact that the Mu spin precession frequency over a time period $\tau_c=\nu^{-1}$ cannot exceed $\omega_{24}$, so that the time-averaged HF field is reduced to $\sim\omega_0/\nu$ due to the motional narrowing effect.  The LF dependence of $P_z(0)$ as a function of $\nu$ is shown in Fig.~\ref{Pz0n}.

Regarding $1/T_{1\mu}$, it is derived from Eq.~(\ref{Acf}) with $Q=1$ via the dynamical susceptibility,
\begin{equation}
\chi(\omega)=\frac{1}{2}\coth\left(\frac{\hbar\omega}{2k_BT}\right)\frac{\chi_{\rm s}}{2\pi}\int C(t)e^{i\omega t}dt,
\end{equation}
\begin{equation}
\chi_{\rm s}=\Delta_{\rm eff}^2/k_BT,
\end{equation}
where the factor $\frac{1}{2}\coth(\frac{\hbar\omega}{2k_BT})$ comes from the fluctuation-dissipation theorem for the thermal average over the canonical ensemble, $\chi_{\rm s}$ is the static susceptibility obeying the Curie law, and $\Delta_{\rm eff}$ ($=\gamma_{\rm av}\sqrt{\langle H(0)^2\rangle}$) is the linewidth determined by the second moment of the fluctuating local field. The general form for the spin relaxation exhibited by a given Mu$^0$ state is expressed in terms of the transitions between four energy levels,
\begin{equation} 
1/T_{1\mu} = \sum_{i,j}\frac{a_{ij}{\rm Im}\:\chi(\omega_{ij})}{\frac{1}{2}\coth(\frac{\hbar\omega_{ij}}{2k_BT})}\approx\Delta_{\rm eff}^2\sum_{i,j}\frac{a_{ij}J(\omega_{ij})}{\omega_{ij}},\label{tone}
\end{equation}
where $J(\omega)$ is the spectral density corresponding to the imaginary part of $\chi(\omega)$,  $\omega_{ij}$ are the relevant Zeeman frequencies of Mu$^0$ with their respective amplitude $a_{ij}$ ($i,j=1$--4).   The Lorentz-type dynamical susceptibility, $\chi(\omega)=\chi_{\rm s}/(1-i\omega/\nu)$ is derived from $C(t)$ to yield 
\begin{equation}
J(\omega)=\frac{\omega/\nu}{1+(\omega/\nu)^2}, \label{BPP}
\end{equation}
which corresponds to the Debye model in dielectric relaxation \cite{Debye:29} (or that introduced by Bloembergen-Purcell-Pound (BPP) or Redfield in NMR \cite{Bloembergen:48}).

\begin{figure}[t]
  \centering
\includegraphics[width=0.95\linewidth]{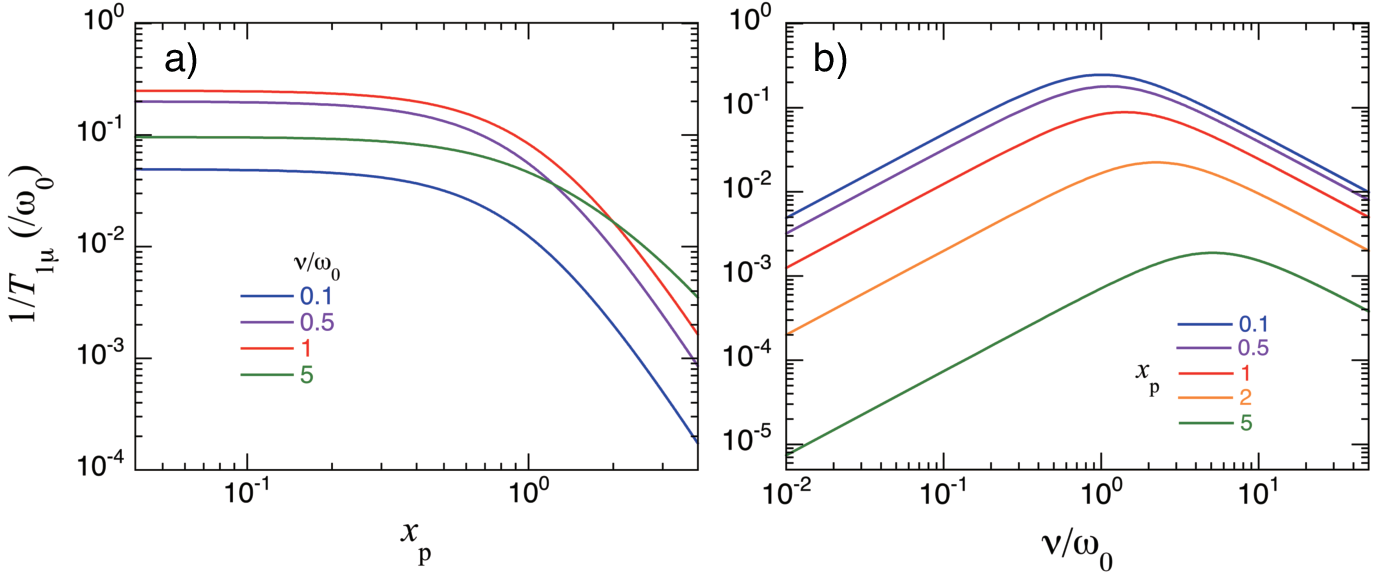}
\caption{(Color online)  (a) LF ($x_{\rm p}$) dependence and (b) fluctuation frequency ($\nu$) dependence of the longitudinal spin relaxation rate ($1/T_{1\mu}$) normalized by the HF frequency ($\omega_0$). ($\Delta_{\rm n}=0$ is assumed.) See Eq.~(\ref{tone24}) for more detail. }
\label{t1scn}
\end{figure}

Now, in the case of spin and/or charge exchange interaction which is dominated by the switching of $\omega_0$, the relevant transition is $\omega_{24}$ with $\Delta_{\rm eff}=\omega_0$, and we have
\begin{equation} 
1/T_{1\mu} \approx\omega_0^2\frac{a_{24}J(\omega_{24})}{\omega_{24}}=\frac{\nu[1-g_z(x_{\rm p})]}{(1+x_{\rm p}^2)+\nu^2/\omega_0^2},\label{tone24}
\end{equation}
where we assumed $a_{24}=1-g_z(x_{\rm p})$.  Eq.~(\ref{tone24}) is further approximated in terms of the relative magnitude of $\nu$ versus $\omega_0$,
\begin{equation} 
1/T_{1\mu} \approx
\left\{
\begin{array}{ll}
\frac{\nu}{2(1+x_{\rm p}^2)},& (\nu\ll\omega_0)\\
\frac{\omega_0^2}{2\nu(1+x_{\rm p}^2)}, &(\nu\gg\omega_0)
\end{array} \label{t1ap}
\right.
\end{equation}
where the case of $\nu\ll\omega_0$ at a lower LF is in agreement with that reported earlier \cite{Nosov:65}. Thus, Eq.~(\ref{tone24}) represents $1/T_{1\mu}$ for arbitrary $\nu$. The relative value of $1/T_{1\mu}$ as a function of $x_{\rm p}$ or $\nu$ is shown in Fig.~\ref{t1scn}. When the HF and NHF fields are comparable (which is often the case for Mu$\dot{R}$s), the effective linewidth may be approximated by $\Delta_{\rm eff}^2\simeq\omega_0^2+\Delta_{\rm n}^2$.  

\subsubsection*{2.2 Diffusive motion of paramagnetic Mu}
\begin{figure}[t]
  \centering
\includegraphics[width=0.65\linewidth]{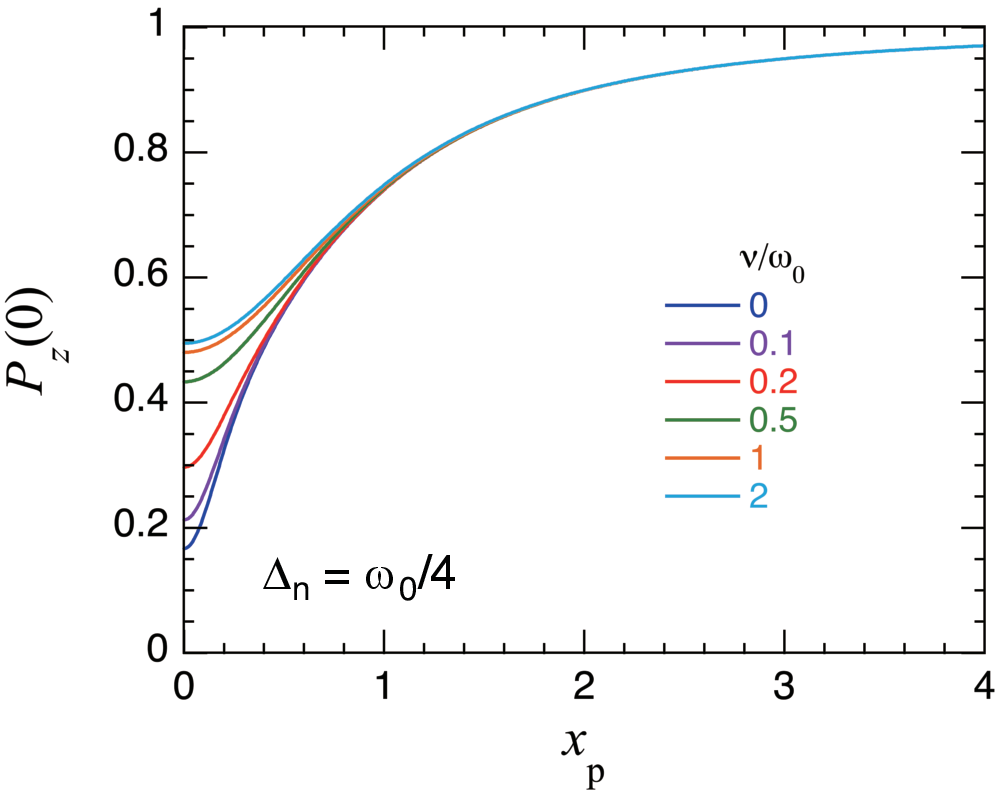}
\caption{(Color online)  LF ($x_{\rm p}$) dependence of the initial muon polarization $P_z(t)$ when only the NHF field ($\Delta_{\rm n}=\omega_0/4$) fluctuates with the rate $\nu$.  $P_z(0)\rightarrow\frac{1}{2}$ with $\nu\gg\Delta_{\rm n}$. }
\label{Pz0no}
\end{figure}

Since Mu$\dot{R}$s found in organic materials are well localized by strong covalent bonds with the constituent atoms of the host, there are few known examples of muons and unpaired electrons moving in unit in the host. On the other hand, such motion is universally observed for Mu$^0$ in inorganic crystals \cite{Kiefl:89,Kadono:90,Kadono:91,Kadono:94,Gxawu:05}. In this case, the fluctuation of the HF field is an indirect effect of the NHF field fluctuating as Mu$^0$ jumps between sites (Fig.~\ref{dymod}(c)), and only the $h_z(x_{\rm n};\nu)$ part is affected by the fluctuation in the LF dependence of the initial polarization. Specifically, $P_z(t)$ is predicted to be described by the following equations:
\begin{equation}
P_z(t;x_{\rm p})\simeq g_z(x_{\rm p};\nu)\exp[-t/T_{1\mu}],  %
\end{equation}
\begin{eqnarray}
g_z(x_{\rm p};\nu)&\simeq& \frac{\frac{1}{2}h_z(x_{\rm n})+x_{\rm p}^2}{1+x_{\rm p}^2},\label{gzpdif}\\
h_z(x_{\rm n};\nu)&\simeq& \frac{\frac{1}{3}+x_{\rm n}^2+\nu^2/\Delta_{\rm n}^2}{1+x_{\rm n}^2+\nu^2/\Delta_{\rm n}^2}.
\end{eqnarray}
Some examples of $P_z(0)$ vs.~LF are shown in Fig.~\ref{Pz0no} for the case of $\Delta_{\rm n}=\omega_0/4$ with various values of $\nu/\omega_0$, where $P_z(0)$ changes from $1/6$ to $1/2$ with increasing $\nu$ around $\Delta_{\rm n}$. Thus, diffusive motion of Mu$^0$ is characterized by the partial recovery of $P_z(0)$ by the motional narrowing, which is largely different from the case for the direct fluctuation of unpaired electron as observed in Fig.~\ref{Pz0n}.  In the Mu$\dot{R}$ case, the anisotropic component of the HF field ($\omega_*$) can be fluctuated by the jump motion, further contributing to $\Delta_{\rm n}$.

\begin{figure}[t]
  \centering
\includegraphics[width=0.95\linewidth]{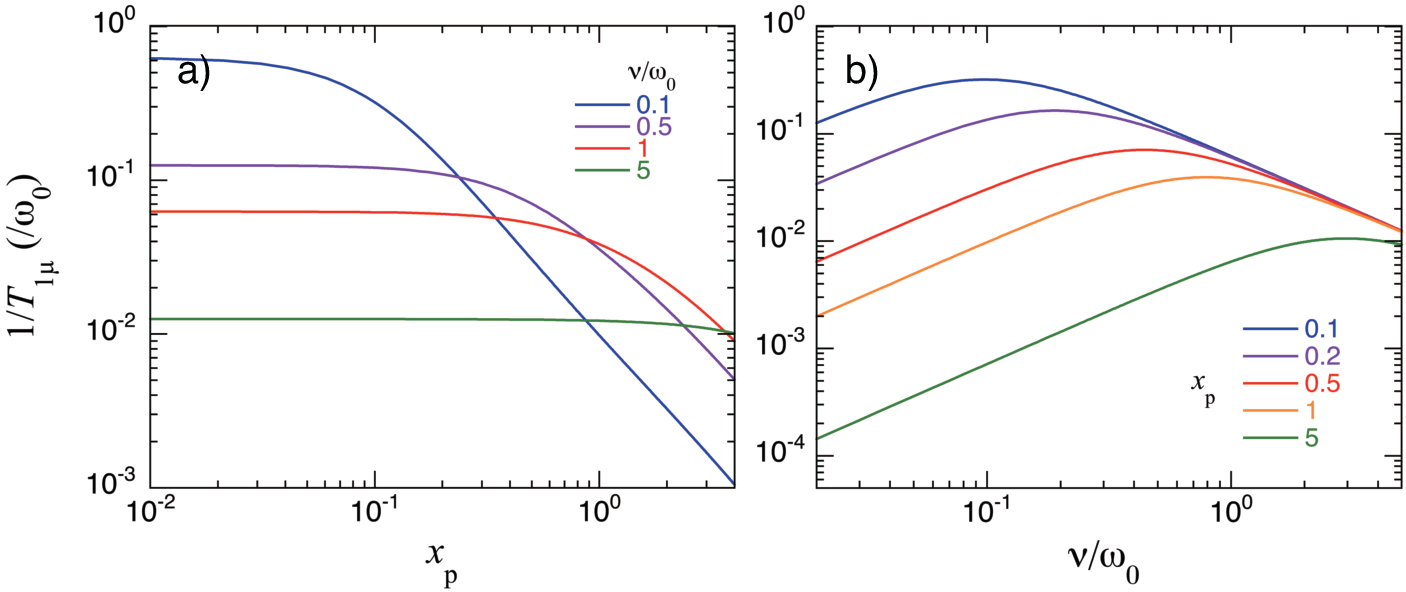}
\caption{(Color online)  (a) LF ($x_{\rm p}$) dependence and (b) fluctuation frequency ($\nu$) dependence of the longitudinal spin relaxation rate ($1/T_{1\mu}$) normalized by the HF frequency ($\omega_0$). ($\Delta_{\rm n}=\omega_0/4$ is assumed.) See Eq.~(\ref{toneno}) for more detail. }
\label{t1no}
\end{figure}

For longitudinal spin relaxation, in contrast to the spin/charge exchange case, all transitions in Eq.~(\ref{tone}) can contribute, so that its LF or fluctuation frequency dependence can only be accurately given by numerical calculations \cite{Kiefl:89,Kadono:90}. However, since relatively low intra-triplet transition frequencies (e.g., $\omega_{12}$) are considered to be the main contributors in the actually observed relaxation, the approximation is given as follows \cite{Kadono:90}:
\begin{eqnarray} 
1/T_{1\mu} &\approx&\frac{\Delta_{\rm n}^2\nu}{\omega_{12}^2+\nu^2}
=\Delta_{\rm n}^2\frac{\nu/\omega_0^2}{x_{12}^2+\nu^2/\omega_0^2},\label{toneno}\\
x_{12}&=&\omega_{12}/\omega_0 \simeq \frac{1}{2}\left[1-(1+x_{\rm p}^2)^{1/2}\right]+x_{\rm p}.
\end{eqnarray}
For the illustration purpose, the LF and $\nu$ dependence of $1/T_{\rm 1\mu}$ calculated by Eq.~(\ref{toneno})  is shown in Fig.~\ref{t1no}.  While the absolute values of $1/T_{\rm 1\mu}$ at a given set of $x_{\rm p}$ and $\nu$ are significantly different between Figs.~\ref{t1scn} and \ref{t1no}, both exhibit qualitatively similar behavior.  Thus, although it is difficult to distinguish these two cases from the information on $1/T_{\rm 1\mu}$ alone, it becomes possible to distinguish them by comparing $P_z(0)$ shown in Figs.~\ref{Pz0n} and \ref{Pz0no} where the dynamic effects are largely different.

\subsubsection{Spin dynamics treated by the partial correlation model ($Q<1$)}

Figure \ref{dymod}(d) illustrates the partial fluctuation in $\Delta_{\rm n}$ due to the diffusive motion of one of the ions adjacent to Mu$^0$. When the average jump rate $\nu_{\rm i}$ of these ions is smaller than $\Delta_{\rm n}$ in such diffusive motion, it is difficult to assume a situation where all neighboring ions jump simultaneously as described in the strong collision model. 

In the case of molecular crystals which have internal degrees of freedom, Mu$\dot{R}$ formed in them may feel the dynamical effects of intramolecular motion and cause longitudinal muon spin relaxation; unpaired electrons associated with Mu$\dot{R}$ are localized at neighboring sites as polarons, so the Mu spin and nuclear spins are subject to electron spin relaxation induced by fluctuations due to molecular motion through the HF field. Such fluctuations are considered to be well approximated by a partial correlation model because the motion of molecules in a crystal lattice is geometrically restricted.
In reality, not only $\Delta_{\rm n}$ but also $\omega_0$ may be modulated by local torsional motion of functional groups and/or side chains in the vicinity of a molecule to which Mu$\dot{R}$ is attached. Indeed, such a situation was inferred in the course of studying the spin relaxation of Mu$\dot{R}$ in the conjugated polymer P3HT \cite{Takeshita:24}.

\begin{figure}[t]
  \centering
\includegraphics[width=0.95\linewidth]{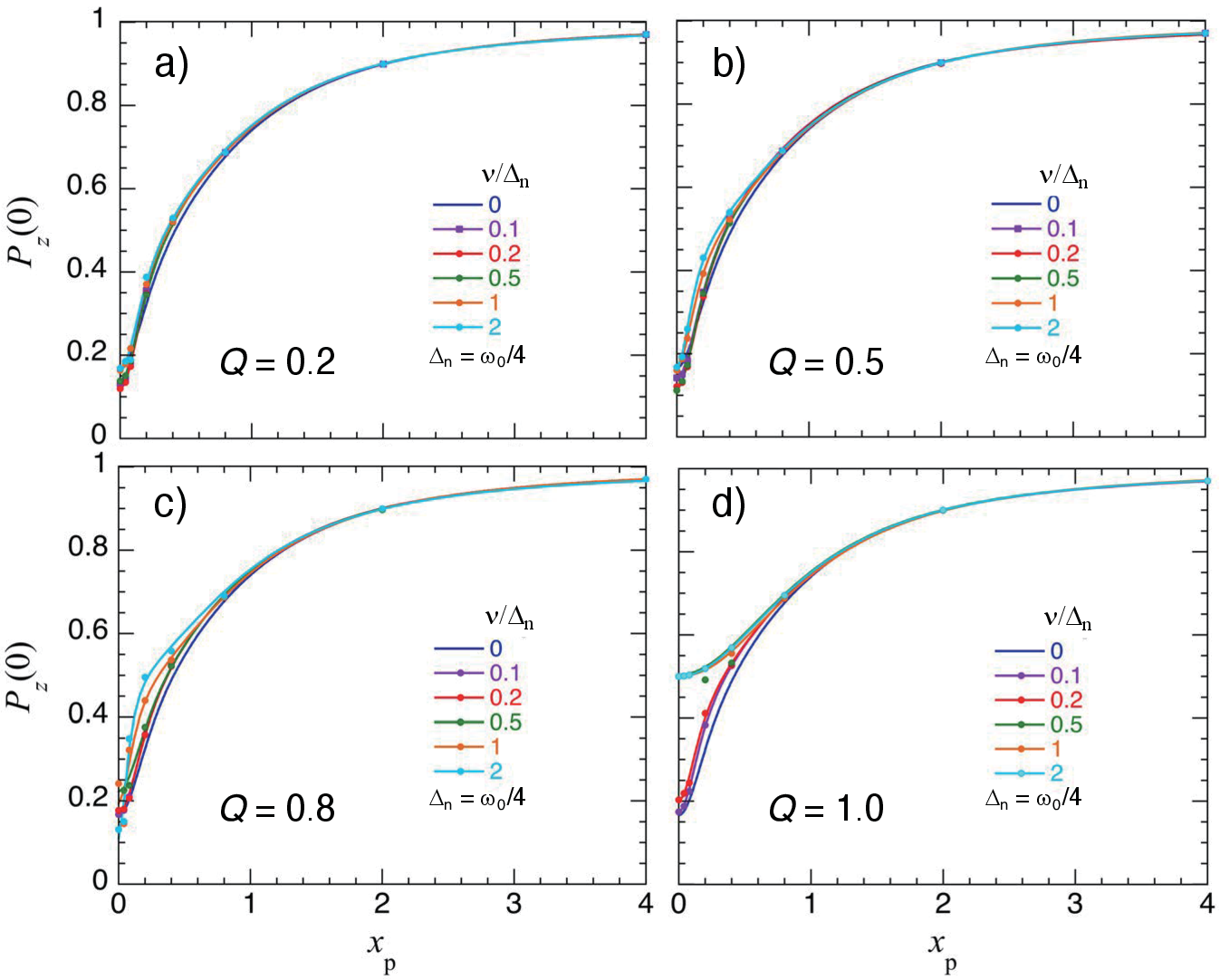}
\caption{(Color online)  Initial muon polarization [$P_z(0)$] versus LF ($x_{\rm p}$) with various fluctuation rate ($\nu_{\rm i}$) for the cases of $Q=0.2$ (a), 0.5 (b), 0.8 (c), and 1.0 (d). Solid lines show the results of curve fit using approximated functions similar to Eq.~(\ref{gzpdif}). }
\label{Pz0pc}
\end{figure}

Because of the difficulty to obtain $P_z(t)$ in analytical form in the partial correlation model except for the case $Q=1$ (or 0), we derived it by Monte Carlo simulations (see Ref.\cite{Ito:24} for simulation conditions). It was presumed that the quasi-static NHF field follows a Gaussian distribution with a linewidth $\Delta_{\rm n}=\omega_0/4$, and that the spin relaxation due to fluctuations of the NHF fields is mainly caused by the intra-triplet transitions which was approximated by the precession with a Zeeman frequency $\gamma_{\rm av}B$.  Fig.~\ref{Pz0pc} shows the LF dependence of the initial polarization $P_z(0)$ for several values of $\nu_{\rm i}$ and $Q$, where  $P_z(0)$ was obtained by averaging  $P_z(t)$ over a time period $0\le t\le3\Delta_{\rm n}^{-1}$ to be compared with the limited time resolution in the conventional experimental conditions: $3\Delta_{\rm n}^{-1}\simeq10$ ns for $\Delta_{\rm n}=2\pi\times50$ MHz.  The solid lines are obtained by curve fitting the values deduced in the simulation for the 8 points of $x_{\rm p}$ ranging from 0 to 4 using a function similar to Eq.~(\ref{gzpdif}). 

 When $Q$ is smaller than unity, the recovery of $P_z(0)$ to $\frac{1}{2}$ with increasing $\nu_{\rm i}$ is sharply suppressed, and the LF dependence of $P_z(0)$ is almost as in the quasi-static case for $Q\lesssim0.5$. In other words, $P_z(0)$ no longer undergoes motional narrowing, which is a distinctive feature compared to the previous strong collision model ($Q=1$). Therefore, even when longitudinal relaxation similar to that in the $Q=1$ case is observed, the cause of the fluctuation can be identified by examining the LF dependence of $P_z(0)$ in detail.

\begin{figure}[t]
  \centering
\includegraphics[width=0.95\linewidth]{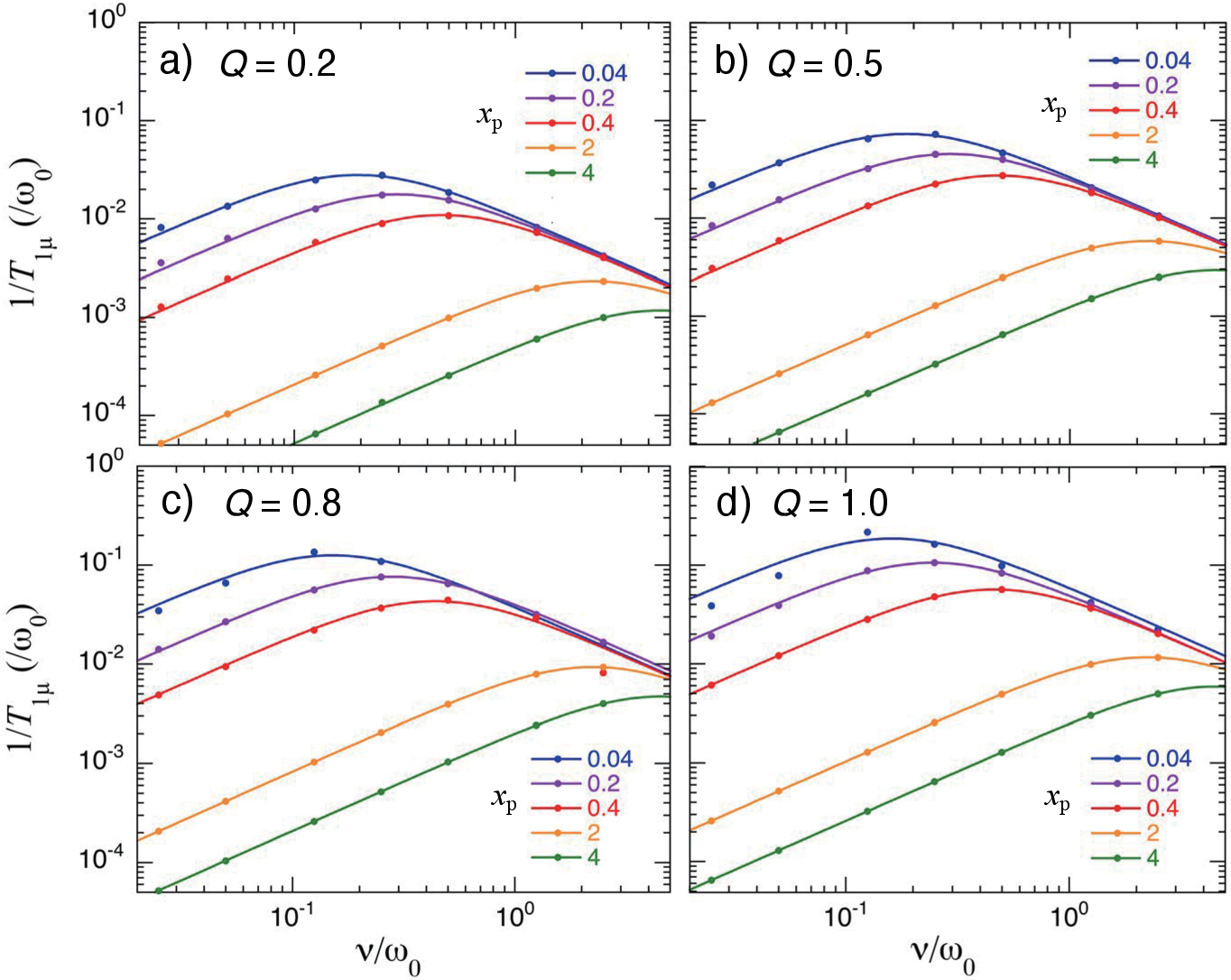}
\caption{(Color online)  Longitudinal relaxation rate ($1/T_{1\mu}$ versus LF ($x_{\rm p}=4x_{\rm n}$)  with various fluctuation rate ($\nu_{\rm i}$) for the cases of $Q=0.2$ (a), 0.5 (b), 0.8 (c), and 1.0 (d). Solid lines show the results of curve fit using the BPP formula (see text). }
\label{T1pc}
\end{figure}

The $\nu_{\rm i}$ dependence of the longitudinal relaxation rate for $t>3\Delta^{-1}$ obtained by the same simulation is shown in Fig.~\ref{T1pc} for several LF ($x_{\rm p}$) and $Q$. Here, the solid line shows the results of curve fitting obtained for 7 points of $\nu_{\rm i}/\omega_0$ ranging from 0.1 to 10, assuming that the spectral density for a given $\nu_{\rm i}$ is described by Eq.~(\ref{BPP}), 
\begin{equation}
1/T_{1\mu}\approx\Delta_{\rm eff}^2\frac{\nu_{\rm i}/\omega_0^2}{x_{\rm p}^2+\nu_{\rm i}^2/\omega_0^2},\label{PCM}
\end{equation}
which shows excellent agreement with simulation data. Although the crude approximation adopted for this simulation does not allow direct comparison between Fig.~\ref{T1pc}(d) and Fig.~\ref{t1no}(b) (corresponding to $Q=1$), they exhibit semi-qualitative agreement with each other.

\begin{figure}[t]
  \centering
\includegraphics[width=0.6\linewidth]{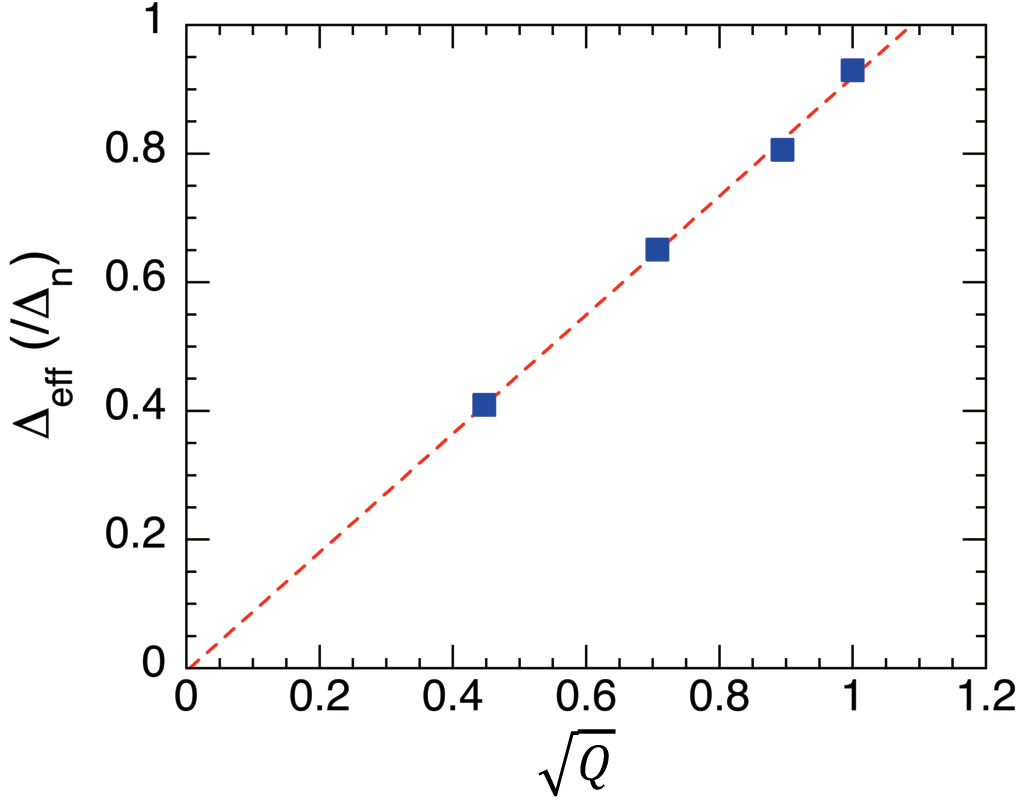}
\caption{(Color online)  The effective NHF field ($\Delta_{\rm eff}$) plotted against $\sqrt{Q}$. The dashed line shows the result of curve fit assuming the linear relation between $\Delta_{\rm eff}$ and $\sqrt{Q}$. }
\label{Dlteff}
\end{figure}

Here, the effect of $Q$ appears in the relative magnitude of $1/T_{1\mu}$ which is determined by $\sqrt{Q}$. As shown in Fig.~\ref{Dlteff}, $\Delta_{\rm eff}$ deduced from these fits plotted against  $\sqrt{Q}$ are on the straight line passing the coordinate origin, indicating that $\Delta_{\rm eff}/\Delta_{\rm n}\simeq\sqrt{Q}$. This is consistent with the expectation in the BPP equation that the effective linewidth $\Delta_{\rm eff}$ under the partial correlation model is given by the relation
\begin{equation}
\Delta_{\rm eff}^2=Q\Delta_{\rm n}^2,\label{Deff}
\end{equation}
as suggested from Eq.~(\ref{Acf}). 

\begin{figure*}[t]
  \centering
\includegraphics[width=0.75\linewidth]{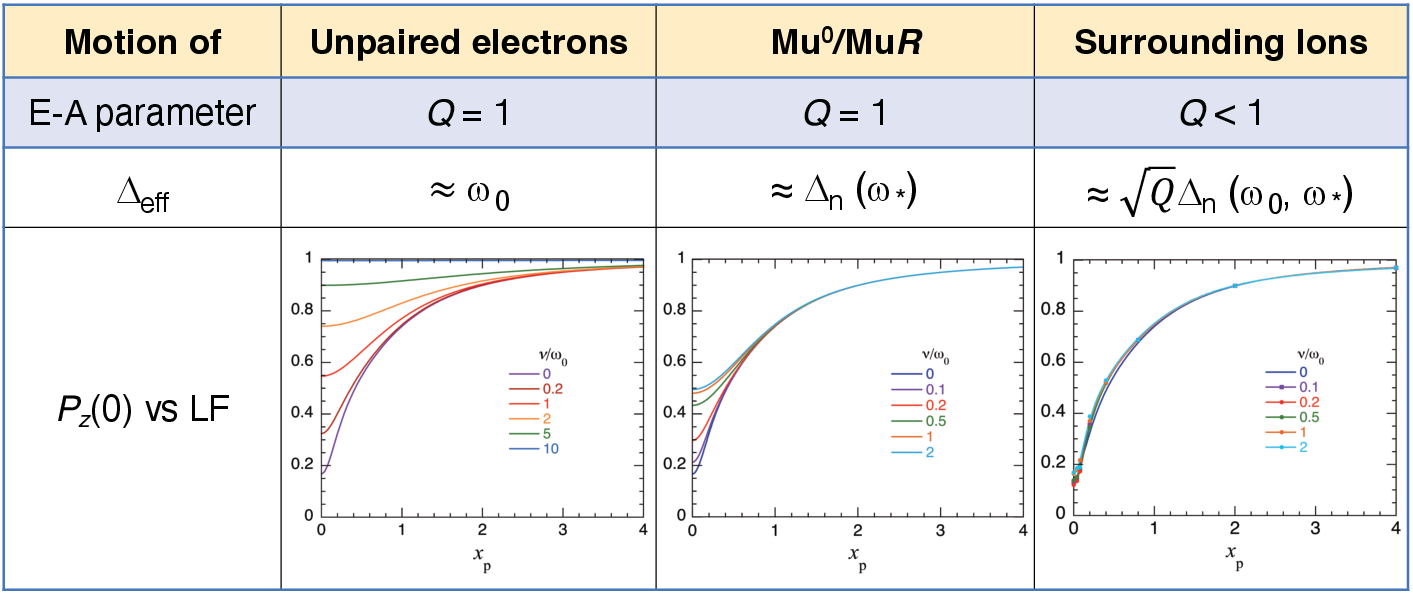}
\caption{(Color online)  Effective linewidth $\Delta_{\rm eff}$ and initial polarization $P_z(0)$ vs.~LF when the dynamical spin relaxation exhibited by Mu/Mu$\dot{R}$ is due to (a) independent motion of unpaired electrons ($Q=1$), (b) Mu/Mu$\dot{R}$ self-diffusion ($Q=1$), and (c) motion of the surrounding ions ($Q<1$).}
\label{PzQsum}
\end{figure*}

It is important to emphasize that analyzing the behavior of $1/T_{1\mu}$ is not enough to distinguish whether the relaxation is due to the self-diffusion of Mu$^0$/Mu$\dot{R}$ or to the dynamics of ions adjacent to Mu$^0$/Mu$\dot{R}$. This is evident by comparing the similarity of Eqs.~(\ref{toneno}) and (\ref{PCM}) and the $\nu$ ($\nu_{\rm i}$) dependence of $1/T_{1\mu}$ expected from them (Fig.~\ref{t1no} vs.~\ref{T1pc}). In ion dynamics, however, the situation of $Q<1$ in the partial correlation model is assumed to be dominant, in which case the behavior of $P_z(0)$ is significantly different from that in the strong collision model ($Q=1$), as is evident from the comparison of Figs.~\ref{Pz0no} and \ref{Pz0pc}.

In the case of spin/charge exchange reactions, the HF field is assumed to have a constant magnitude and the residual part of the autocorrelations upon fluctuation are induced by spatially restricted motion of molecules. On the other hand, ${\bm H}(t)$ that describes the NHF fields has a static Gaussian distribution and fluctuates according to the same probability distribution at any given time $t$.  Therefore, it is not necessarily obvious whether our model can be directly applied to the case of fluctuating HF fields, and further investigation may be required to find a more appropriate model.

However, the behavior of $P_z(0)$ and $1/T_{1\mu}$ for the partially fluctuating HF fields is expected to be qualitatively the same as the results obtained by applying the partial correlation model to the NHF field. Specifically, the recovery of $P_z(0)$ with increasing $\nu_{\rm i}$, as seen in Fig.~\ref{Pz0n}, is expected to be strongly suppressed for $Q<1$. Moreover, $\Delta_{\rm eff}$ would be smaller in proportion to $\sqrt{Q}$ with decreasing $Q$: a relation similar to Eq.~(\ref{Deff}) is expected (e.g., $\Delta_{\rm eff}^2\approx Q\omega_0^2$). Therefore, it would be possible to infer the cause of the fluctuations depending on whether these behaviors are observed experimentally.  These characteristic  features for the three respective cases are summarized in Fig.~\ref{PzQsum}.

\section{Discussion}
There have been reports on the observation of spin relaxation due to the motion of unpaired electrons on a quasi-1D molecular chain by $\mu$SR for several conducting polymers. However, a detailed examination of these reports suggests potential shortage of information for judging the validity of such an interpretation. 

\begin{figure}[b]
  \centering
\includegraphics[width=0.95\linewidth]{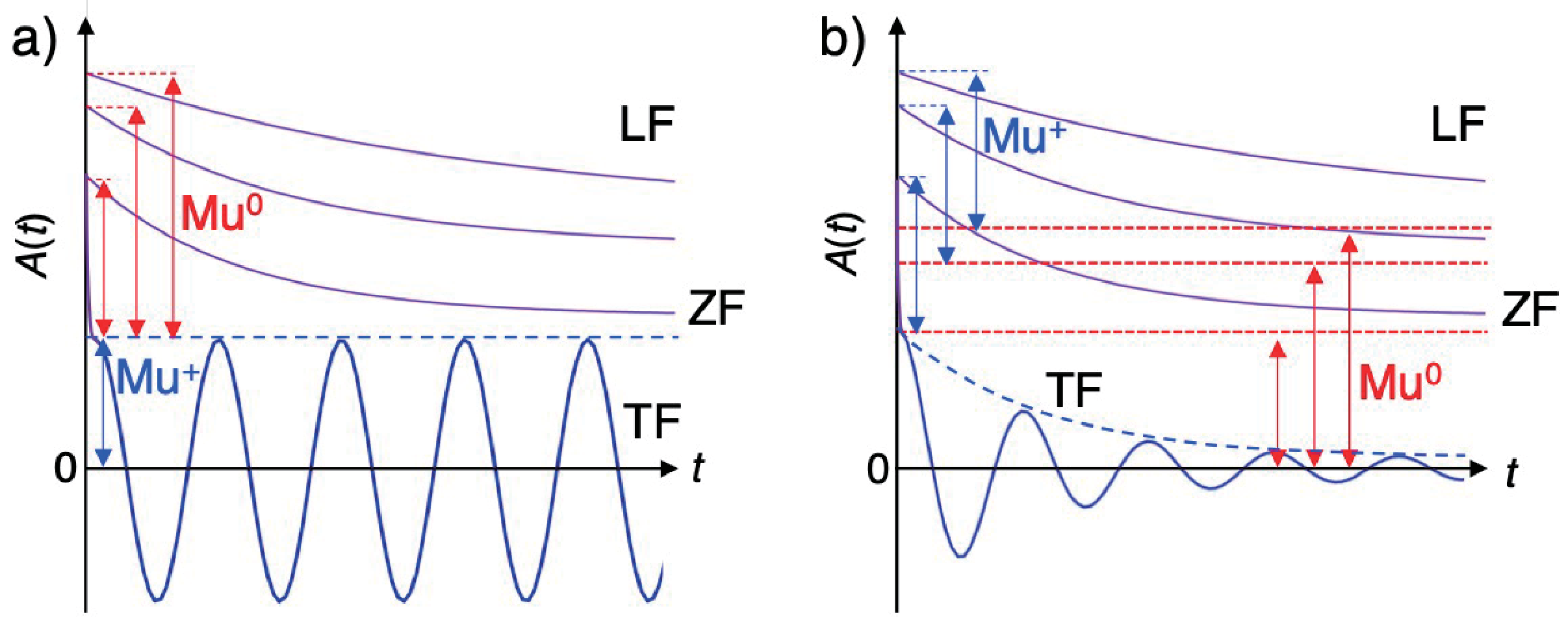}
\caption{(Color online)  Magnetic field dependence of $\mu$SR time spectra $A(t)$ when Mu$^+$ (diamagnetic Mu) and Mu$^0$ (Mu$\dot{R}$) coexist in comparable yields.  When the spin relaxation under ZF/LF is mainly due to either Mu$^0$ (a) or Mu$^+$ (b), the two can be distinguished in the TF-$\mu$SR spectra.}
\label{LZTF}
\end{figure}

One of the issues is that when there are multiple Mu states, it is not always obvious which state the observed spin relaxation originates from. In the earlier reports on PANI and P3HT, both diamagnetic Mu and Mu$\dot{R}$ have been suggested to coexist, where the spin relaxation is attributed to the diamagnetic Mu (which is implicitly assumed to originate from another Mu$\dot{R}$ state close to the motional narrowing limit) \cite{Pratt:97,Blundell:02,Pratt:04,Risdiana:10a,Risdiana:10b}. However, the relative yields of these components and their respective contributions to the relaxation rate under LF are not clear from the reported information. A similar situation is found in the seminal case of PA, especially for {\it trans}-PA (see below) \cite{Nagamine:84,Ishida:85}. In any case, since the effective gyromagnetic ratio of Mu$\dot{R}$ is two orders of magnitude larger than that of diamagnetic Mu and is sensitive to small amplitude fluctuations, there is a possibility that Mu$\dot{R}$ shows spin relaxation due to other causes such as local motion of molecular chains.  This makes it crucial to be able to quantitatively separate the contribution of these two components to the spin relaxation.
As shown schematically in Fig.~\ref{LZTF} for the two typical cases, while it is difficult to determine which one is exhibiting spin relaxation only from the LF dependence of the initial asymmetry and/or the relaxation rate when Mu$^+ $ and Mu$\dot{R}$ coexist, it may be possible to distinguish between the two by measuring $\mu$SR spectra under a weak transverse field (TF) where the signal from the quasi-static Mu$\dot{R}$ is quenched.

Let us now examine the case of P3HT in detail as a typical example. It was argued in earlier reports that the observed $1/T_{1\mu}$ was due to the 1D motion of unpaired electrons along the polythiophene chains \cite{Risdiana:10a,Risdiana:10b}. The primary basis for this claim is that the LF dependence of $1/T_{1\mu}$ at low temperatures follows $J(\omega)/\omega\propto\omega^{-1/2}$ as theoretically predicted for 1D jump motion of electrons, where it is implicitly assumed that the spin relaxation originates from diamagnetic Mu (corresponding to Fig.~\ref{LZTF}(b)).

However, it has been demonstrated in the recent $\mu$SR study that the spin relaxation rate exhibited by diamagnetic Mu in TF-$\mu$SR spectra is significantly smaller than that under ZF (corresponding to Fig.~\ref{LZTF}(a)), indicating that the ZF/LF spectra mainly reflect spin relaxation in Mu$\dot{R}$ \cite{Takeshita:24}. Moreover, the preceding NMR study reported the 1D carrier diffusion rate $D_\parallel\simeq5.3\times10^{14}$ s$^{-1}$ \cite{Mabboux:95}. Assuming that the charge exchange rate is approximately given as $\nu\simeq D_\parallel$, it is far greater than $\omega_0$ ($=2\pi\times 220$ MHz $\simeq10^9$ s$^{-1}$).  Therefore, when the unpaired electron associated with Mu$\dot{R}$ induces charge exchange reaction at such high frequencies ($\nu\gg\omega_0$), its initial asymmetry ($A(0)$, $\propto P_z(0)$) would be of full value corresponding to $P_z(0)=1$ regardless of LF (see Fig.~\ref{Pz0n} or \ref{PzQsum}(a)). A similar behavior of $P_z(0)$ is also predicted in the more specific model that considers expansion of a domain on the polymer chain where the 1D jump motion occurs with time after Mu addition \cite{Risch:92}. These predictions are not consistent with the observation that $A(0)$ for the Mu$\dot{R}$ component exhibits strong LF dependence corresponding to $\nu\lesssim\omega_0$ \cite{Takeshita:24}. 

As another issue, they report a large deviation of $1/T_{1\mu}$ from this $\omega^{-1/2}$ behavior around room temperature, which they attribute to the onset of three-dimensional (3D) diffusion of carriers \cite{Risdiana:10a,Risdiana:10b}. However, this is not consistent with the implication from NMR that the 1D--3D anisotropy remains large ($D_\parallel/D_\perp>10^6$) in the relevant temperature range \cite{Mabboux:95}.  

As mentioned above, it has recently been shown that $1/T_{1\mu}$ originates from Mu$\dot{R}$ \cite{Takeshita:24}. In addition, the detailed analysis based on the Havriliak-Negami (HN) function \cite{Havriliak:67} for the susceptibility,
\begin{equation}
\chi(\omega)=\chi_{\rm s}\frac{1}{[1-i(\omega/\tilde{\nu})^\delta]^\gamma},\label{HN}
\end{equation}
incorporated into $1/T_{1\mu}$ (where $0\le\delta,\gamma\le1$) yielded the mean fluctuation rate $\tilde{\nu}\simeq10^9$--$10^{10}$ s$^{-1}$. (Note that Eq.~(\ref{HN}) is a generalization of the Lorentz-type $\chi(\omega)$ with $\delta=\gamma=1$.) The magnitude of $\tilde{\nu}$ suggests that the fluctuation is induced by the local molecular motion \cite{Takeshita:24}, which is also consistent with the conclusions brought by $^{13}$C-NMR studies \cite{Yazawa:10}. Furthermore, the value of $\Delta_{\rm eff}$ obtained from the $1/T_{1\mu}$ analysis is $\sim$40\% of $\Delta_{\rm n}$ at lower temperatures, which is interpreted as corresponding to the situation where $Q\approx0.16$. This is also in line with the quasi-static behavior exhibited by the LF dependence of $P_z(0)$ expected for $Q<1$ (see  Fig.~\ref{Pz0pc}(a) or \ref{PzQsum}(c)). The observed change in the LF dependence of $1/T_{1\mu}$ with temperature (the reported `deviation' from the $\omega^{-1/2}$ dependence in the earlier literature) is then attributed to the change in $\Delta_{\rm eff}$ (that also accompanies a slight change in $\delta$ and $\gamma$) due to the successive activation of molecular motions from twisting of hexyl chains to bending of thiophene rings \cite{Yazawa:10}. Therefore, it is concluded that the origin of the fluctuations observed in $\mu$SR is neither the 1D motion of the unpaired electron nor the diffusion of Mu$\dot{R}$, but the local motion of the molecules around Mu$\dot{R}$. Needless to mention that, with this interpretation, there is also no contradiction between $\mu$SR and NMR for 1D--3D anisotropy in the carrier diffusion rate.

A similar issue exists in the $\mu$SR studies on PANI \cite{Pratt:97}: the reported $A(0)$ in the LF-$\mu$SR time spectra at 6 K shows a strong LF dependence, indicating the existence of Mu$\dot{R}$. On the other hand, it is assumed that the observed $1/T_{1\mu}$ originates from a diamagnetic Mu state, mainly on the basis that the time-dependent lineshape is better reproduced by that predicted in Ref.~\cite{Risch:92} than the simple exponential decay, and concluded that $D_\parallel$ is about $10^{14}$ s$^{-1}$.  However, the possibility that PANI is in the same situation as P3HT cannot be ruled out at this stage. To further investigate the validity of this interpretation, TF-$\mu$SR data at relevant temperatures are needed to quantitatively evaluate the diamagnetic Mu contribution.

As a hint for understanding why such a contradictory interpretation is possible, we should recall a general feature that $\nu$ (or $\nu_{\rm i}$) is not uniquely deduced from $1/T_{1\mu}$.  For example, approximating Eq.~(\ref{HN}) in the two limit regions of $\tilde{\nu}$ relative to LF yields
\begin{equation}
1/T_{1\mu}\propto\frac{J(\omega)}{\omega}
\simeq
\left\{
\begin{array}{ll}
\frac{\tilde{\nu}^{\gamma\delta}}{\omega^{\gamma\delta+1}} &(\tilde{\nu}\ll\omega)\\
\frac{\omega^{\delta-1}}{\tilde{\nu}^\delta} & (\tilde{\nu}\gg\omega) 
\end{array} \label{jw}
\right.
\end{equation}
which indicates two different values of $\tilde{\nu}$ for $\tilde{\nu}>\omega$ and $\tilde{\nu}<\omega$ can give the same value for $1/T_{1\mu}$. Therefore, to determine at which side of  the $T_1$-minimum the corresponding $\nu$ is situated, it is necessary to consider not only the $1/T_{1\mu}$ information but also the LF dependence of $P_z(0)$, as shown in Fig.~\ref{PzQsum}.

Finally, we also point out the remaining issues regarding the earlier $\mu$SR studies on polyacetylene, where $1/T_{1\mu}$ in {\it trans}-PA is attributed to the soliton-like 1D motion of unpaired electrons on the conjugated molecular chains \cite{Nagamine:84,Ishida:85}. Reportedly, in {\it cis}-PA where the unpaired electron is presumed to be localized (i.e.~the polaron state), $A(0)$ is smaller than $A_0/2$ at zero field (where $A_0$ is the full asymmetry corresponding to 100\% $\mu^+$ polarization) and recovers with increasing LF, suggesting that the Mu$\dot{R}$ feels a quasi-static HF and NHF fields. On the other hand, in {\it trans}-PA, $A(0)$ takes a value greater than $A_0/2$ at ZF and recovers to $A_0$ with LF. The $\mu$SR spectra under a weak TF ($\simeq2$ mT, recently retrieved from TRIUMF data archive) suggests that the ZF-$\mu$SR spectra mainly consist of a diamagnetic Mu state, not the spin-triplet state of Mu$\dot{R}$ \cite{Ishida:84}. Therefore, it is strongly suggested that both Mu$\dot{R}$ and a diamagnetic Mu coexist in comparable yield, where the former exhibits negligible initial asymmetry at ZF common to {\it cis}-PA. This situation calls into question the assumption in the interpretation at the time that all implanted muons were in an apparent diamagnetic Mu state reflecting the fast soliton motion of unpaired electrons with $\nu\approx10^{12}$ s$^{-1}$ ($\gg\omega_0$). Thus, further studies are needed to determine the origin of the observed muon spin relaxation.

\section{Summary and Conclusion}

We have clarified how the spin dynamics described by the partial correlation model is reflected in $P_z(0)$ and $1/T_{1\mu}$ for paramagnetic Mu states observed in nonmetallic host materials. In this dynamical model, we can distinguish among the jumping motions of unpaired electrons, self-diffusive motions of paramagnetic Mu, and the local motions of surrounding ions as the origins of the observed fluctuations. Based on this model, we have investigated the interpretations of the unpaired electron motion reported for the muonated radical states in several conducting polymers, and found that they need reconsideration for the coherent understanding of $P_z(0)$ and $1/T_{1\mu}$. This model has been successfully applied to the case of muonated radicals in P3HT to identify the origin of spin relaxation, and is expected to be useful for future studies of various local dynamics associated with paramagnetic Mu states.

\section*{Acknowledgment}
We thank S. Takeshita, F. Pratt, and K. Ishida for helpful discussions during the preparation of the manuscript.  This work was partially supported by JSPS KAKENHI (Grant Nos. 23K11707 and 24H00477) and the MEXT Program: Data Creation and Utilization Type Material Research and Development Project (Grant No. JPMXP1122683430).

\let\doi\relax
\input{Mpara.bbl}
\end{document}

%% file: Mpara.bbl
%